\documentstyle[amssymb,aps]{revtex}

\begin{document}
\title{Interfering with decay of a single photon in microwave cavities through
single photon quantum non-demolition scheme}
\date{\today}
\author{J. G. Peixoto de Faria$^{1}$\thanks{%
Electronic address: jgfaria@lcc.ufmg.br}, P. Nussenzveig$^{2}$, A. F. R. de
Toledo Piza$^{2}$ and M. C. Nemes$^{1}$}
\address{$^1$Departamento de F\'{\i}sica, ICEx, Universidade\\
Federal de Minas Gerais, \\
CP 702, CEP 30123-970 Belo Horizonte, M.G., Brazil \\
$^2$Instituto de F\'{\i}sica, Universidade de S\~ao Paulo, CP 66318,\\
05315-970 S\~ao Paulo, S.P., Brazil}
\maketitle

\begin{abstract}
The decay of a single photon state in a microwave cavity is shown to be
retarded by interaction with a resonant two-level atom in the experimental
setup recently developed by Nogues {\it et al.} [see NOGUES G.,
RAUSCHENBEUTEL A., OSNAGHI S., BRUNE M., RAIMOND J. M. and HAROCHE S., {\it %
Nature}, {\bf 400} (1999) 239]. The effect may be interpreted in terms of
the temporary removal of the photon from the cavity thereby protecting it
from the effects of the environment to which the cavity is coupled.
Realistic parameters lead to a significative increase of the survival
probability of the photon subsequently to the monitoring interaction.
\end{abstract}

\vskip1truecm

A new non-destructive scheme to measure the presence of a single photon in
QED cavity (SP-QND, for single photon quantum non-demolition scheme) has
been recently implemented and described\cite{nogues} by Nogues {\it et al.}.
This scheme differs from the previously available procedure\cite
{brune2,brune1}, using dispersive coupling of the cavity mode to an atomic
probe, in that the probe atom interacts resonantly with the relevant mode,
the interaction time being adjusted so that the system executes a full Rabi
cycle while in contact. In this way the atom ends up in its initial state
after a full coherent cycle of photon absorption and re-emission. This
clearly violates one of the basic requirements for a ``non-demolition''
measurement of the photon number, namely that the measured quantity should
not be affected by the measurement. In this experiment, however, the changes
in photon number are reversible, so that the final state contains ideally
the same photon number as the initial state.

Here we explore a rather subtle effect of the coherent cycle of photon
absorption and re-emission on the decay of the initial single photon
excitation. This decay is due to the unavoidable coupling of the cavity mode
to its environment, and can be infered e.g. from the change in the
probability of having the one photon state at two different times. This can
be done, in principle, using the scheme of Nogues {\it et al.}. As shown
below, if an additional atom executes a full Rabi cycle in the time lapse
between these two measurements, the latter probability is predicted to
increase relatively to the value it would have in the absence of this extra
interaction. A simple interpretation of this result follows if we note that
during the resonant atom-field interaction the photon is actually absorbed
by the atom and is thereby decoupled from the decay process due to the
coupling of the relevant mode to its environment. Since the free lifetime of
the atomic excitation is much larger than the photon lifetime in the cavity
and the transit time of the atom, the effects of environment degrees of
freedom on the atomic excitation can in practice be ignored.

Following ref. \cite{nogues}, we consider three Rydberg circular levels $i$, 
$g$ and $e$ of the probe atom. The Ramsey zones are tuned to the frequency
of the $i\leftrightarrow g$ transition, while the high-Q cavity is tuned to
the (sufficiently different) frequency of the $g\leftrightarrow e$
transition. The dressed dynamics of the atom in the leaky high-Q cavity is
described in terms of the master equation, written in the interaction
picture,

\begin{eqnarray}
\frac{d}{dt}\widehat{\rho }\left( t\right) &=&-i\frac{\Omega }{2}\left[ 
\widehat{a}^{\dagger }\widehat{\sigma }_{-}+\widehat{a}\widehat{\sigma }_{+},%
\widehat{\rho }\left( t\right) \right]  \label{meq} \\
&&+k\left[ 2\widehat{a}\widehat{\rho }\left( t\right) \widehat{a}^{\dagger }-%
\widehat{a}^{\dagger }\widehat{a}\widehat{\rho }\left( t\right) -\widehat{%
\rho }\left( t\right) \widehat{a}^{\dagger }\widehat{a}\right] ,  \nonumber
\end{eqnarray}

\noindent where $\widehat{\rho}$ is the reduced density operator of the atom
plus field system, $\Omega$ is the vaccuum Rabi frequency, the $\widehat{%
\sigma}_i$ are Pauli matrices defined as $\widehat{\sigma}_ {z}\equiv \left|
e\right\rangle\left\langle e\right| -\left| g\right \rangle \left\langle
g\right| $, $\widehat{\sigma }_{-}=\widehat{ \sigma }_{+}^{\dagger }\equiv
\left|g\right\rangle \left\langle e\right| $ and $k$ is the damping constant
related to the Q-value of the microwave cavity. The reduced density operator
can be expanded as

\begin{eqnarray}
\widehat{\rho }\left( t\right) &=&\widehat{\rho }_{ee}\left( t\right)
\otimes \left| e\right\rangle \left\langle e\right| +\widehat{\rho }
_{gg}\left( t\right) \otimes \left| g\right\rangle \left\langle g\right| + 
\widehat{\rho }_{ii}\left( t\right) \otimes \left| i\right\rangle
\left\langle i\right|  \label{rho-1} \\
&&+\left[ \widehat{\rho }_{ge}\left( t\right) \otimes \left| g\right\rangle
\left\langle e\right| +\widehat{\rho }_{ie}\left( t\right) \otimes \left|
i\right\rangle \left\langle e\right| +\widehat{\rho }_{ig}\left( t\right)
\otimes \left| i\right\rangle \left\langle g\right| +{\rm h.\;c.}\right] , 
\nonumber
\end{eqnarray}

\noindent A rather unwieldy set of coupled equations for the operators $%
\widehat{\rho }_{ee}\left( t\right) \equiv \left\langle e\right| \widehat{%
\rho }\left( t\right) \left| e\right\rangle $, etc, can be obtained from eq.
(\ref{meq}). They can however be handled by solving the equations of motion
for their matrix elements in the restricted subspace of zero and one
excitations. As an example, the matrix element $\rho _{e0,e0}\left( t\right)
\equiv \left\langle 0\right| \widehat{\rho }_{ee}\left( t\right) \left|
0\right\rangle =\left\langle e0\right| \widehat{\rho }\left( t\right) \left|
e0\right\rangle $ obeys the equation of motion

\[
\frac{d}{dt}\rho _{e0,e0}\left( t\right) =-i\frac{\Omega }{2}\left[ \rho
_{g1,e0}\left( t\right) -\rho _{e0,g1}\left( t\right) \right] +2k\rho
_{e1,e1}\left( t\right) , 
\]

\noindent which is coupled to the equations of motion for $\rho_{g1, e0}$, $%
\rho_{e0,g1}$ and $\rho _{e1,e1}$. Since the state $e1$ contains two
excitations, the last term in the above equation is neglected. A
straightforward formal solution of the resulting equations leads to, using
again the matrix element $\rho_{e0,e0} \left(t\right)$ as an example,

\[
\rho _{e0,e0}\left( t\right) =e^{-kt}\left[ \alpha \left( t\right) +\beta
\left( t\right) \right] , 
\]

\noindent where $\alpha\left(t\right)$ and $\beta\left(t\right)$ are given by

\begin{equation}
\alpha \left( t\right) =\frac{1}{2}\left[ \rho _{e0,e0}\left( 0\right) +\rho
_{g1,g1}\left( 0\right) \right] +\frac{k}{\Gamma }\left[ \beta _{1}\left(
e^{\Gamma t}-1\right) -\beta _{2}\left( e^{-\Gamma t}-1\right) \right]
\label{alfa}
\end{equation}

\noindent and

\begin{equation}
\beta \left( t\right) =\beta _{1}e^{\Gamma t}+\beta _{2}e^{-\Gamma t}.
\label{beta}
\end{equation}

\noindent The constants $\beta_{1}$ and $\beta_{2}$ are given by

\[
\beta _{1}=\frac{1}{4\Gamma }\left[ \left( \Gamma +k\right) \rho
_{e0,e0}\left( 0\right) +\left( k-\Gamma \right) \rho _{g1,g1}\left(
0\right) \right] +\frac{i\Omega }{4\Gamma }\left[ \rho _{e0,g1}\left(
0\right) -\rho _{g1,e0}\left( 0\right) \right] 
\]

\noindent and

\[
\beta _{2}=\frac{1}{4\Gamma }\left[ \left( \Gamma -k\right) \rho
_{e0,e0}\left( 0\right) -\left( k+\Gamma \right) \rho _{g1,g1}\left(
0\right) \right] -\frac{i\Omega }{4\Gamma }\left[ \rho _{e0,g1}\left(
0\right) -\rho _{g1,e0}\left( 0\right) \right] . 
\]

\noindent The parameter $\Gamma $ appearing in these expressions is formally
given as $\Gamma =\sqrt{k^{2}-\Omega ^{2}}$. In the experimentally relevant
regime of sub-critical damping ($k\sim 10^{3}{\rm s}^{-1}$ and $\Omega /2\pi
\sim 50{\rm kHz}$) we use $\Gamma =i\sqrt{\Omega ^{2}-k^{2}}\equiv i\Omega
^{\prime }$. Thus the frequency with wich the atom and the field exchange
excitations is shifted by the damping to a value $\Omega ^{\prime }$
slightly lower than $\Omega $. For the initial state

\[
\widehat{\rho }\left( 0\right) =\left| g\right\rangle \left\langle g\right|
\otimes \left| 1\right\rangle \left\langle 1\right|, \nonumber
\]

\noindent the matrix element $\rho _{e0,e0}\left( t\right)$ appears as

\[
\rho _{e0,e0}\left( t\right) =\frac{e^{-kt}}{2}\left\{ 1-\frac{1}{\Omega
^{\prime 2}}\left[ \left( k^{2}+\Omega ^{\prime 2}\right) \cos \Omega
^{\prime }t-k^{2}\right] \right\} . 
\]

\noindent After an effective interaction time $t_{i}=2\pi /\Omega ^{\prime }$%
, the probability of finding the atom in the state $e$ vanishes.

\vspace{.3cm}

\noindent {\bf Monitoring the state of the field.} Let an atom prepared in
state $g$ be sent through the apparatus consisting of the cavity, tuned to
the $g\leftrightarrow e$ transition, set between a pair of Ramsey zones
tuned to the $i\leftrightarrow g$ transition so that the effect of each of
them on the state of the atom is

\begin{eqnarray}
\left| g\right\rangle &\rightarrow &\frac{1}{\sqrt{2}}\left( \left|
i\right\rangle +\left| g\right\rangle \right) ,  \nonumber \\
\left| i\right\rangle &\rightarrow &\frac{1}{\sqrt{2}}\left( \left|
i\right\rangle -\left| g\right\rangle \right) .  \label{g-pi/2}
\end{eqnarray}

\noindent Assuming that the initial state of the field in the high-Q cavity
is given by

\[
\widehat{\rho }_{F}\left( 0\right) =P_{1}\left| 1\right\rangle \left\langle
1\right| +P_{0}\left| 0\right\rangle \left\langle 0\right| , \nonumber
\]

\noindent with $P_{1}+P_{0}=1$, the initial state of the composite system
will be

\[
\widehat{\rho }\left( 0\right) =\frac{1}{2}\left( \left| i\right\rangle
+\left| g\right\rangle \right) \left( \left\langle i\right| +\left\langle
g\right| \right) \otimes \left( P_{1}\left| 1\right\rangle \left\langle
1\right| +P_{0}\left| 0\right\rangle \left\langle 0\right| \right) . 
\]

\noindent This initial condition gives for the relevant matrix elements

\[
\rho _{g1,g1}\left( t\right) =\frac{P_{1}}{4}e^{-kt}\left\{ 1-\frac{1}{%
\Omega ^{\prime 2}}\left[ \left( k^{2}-\Omega ^{\prime 2}\right) \cos \Omega
^{\prime }t+2\Omega ^{\prime }k\sin \Omega ^{\prime }t-k^{2}\right] \right\}
, 
\]

\[
\rho _{e0,e0}\left( t\right) =\frac{P_{1}}{4}e^{-kt}\left\{ 1-\frac{1}{%
\Omega ^{\prime 2}}\left[ \left( k^{2}+\Omega ^{\prime 2}\right) \cos \Omega
^{\prime }t-k^{2}\right] \right\} , 
\]

\[
\rho _{e0,g1}\left( t\right) =-i\frac{\Omega P_{1}}{4\Omega ^{\prime 2}}%
e^{-kt}\left( k\cos \Omega ^{\prime }t+\Omega ^{\prime }\sin \Omega ^{\prime
}t-k\right) , 
\]

\[
\rho _{g0,g0}\left( t\right) =\frac{1}{2}\left\{ 1-P_{1}e^{-kt}\left[ 1+%
\frac{k}{\Omega ^{\prime 2}}\left( k-k\cos \Omega ^{\prime }t-\Omega
^{\prime }\sin \Omega ^{\prime }t\right) \right] \right\} , 
\]

\[
\rho _{e0,i1}\left( t\right) =-i\frac{P_{1}}{2}e^{-3kt/2}\sin \left( \frac{%
\Omega ^{\prime }}{2}t\right) , 
\]

\[
\rho _{g1,i1}\left( t\right) =\frac{P_{1}}{2\Omega ^{\prime }}e^{-3kt/2}%
\left[ \Omega ^{\prime }\cos \left( \frac{\Omega ^{\prime }}{2}t\right)
-k\sin \left( \frac{\Omega ^{\prime }}{2}t\right) \right] , 
\]

\begin{eqnarray*}
\rho _{g0,i0}\left( t\right) &=&\frac{1-P_{1}}{2}+i\frac{k}{\Omega ^{\prime }%
}P_{1}\left[ \left( k+i\Omega ^{\prime }\right) \frac{e^{-1/2\left(
3k+i\Omega ^{\prime }\right) t}-1}{3k+i\Omega ^{\prime }}\right. \\
&&\left. -\left( k-i\Omega ^{\prime }\right) \frac{e^{-1/2\left( 3k-i\Omega
^{\prime }\right) t}-1}{3k-i\Omega ^{\prime }}\right] ,
\end{eqnarray*}

\[
\rho _{g0,g1}\left( t\right) =\rho _{g0,e0}\left( t\right) =\rho
_{g0,e1}\left( t\right) =\rho _{g1,i0}\left( t\right) =\rho _{e0,i0}\left(
t\right) =\rho _{g0,i1}\left( t\right) =0. 
\]

\noindent The time evolution of the operator $\widehat{\rho}_{ii}$ is also
needed. It can be obtained algebraically as

\[
\widehat{\rho }_{ii}\left( t\right) =\frac{1}{2}\left[ P_{1}e^{-2kt}\left|
1\right\rangle \left\langle 1\right| +\left( 1-P_{1}e^{-2kt}\right) \left|
0\right\rangle \left\langle 0\right| \right]. 
\]

\noindent When the effective interaction time coincides with a full shifted
Rabi cycle $2\pi/\Omega^{\prime}$ the matrix elements associated with level $%
e$ vanish. As a result, when the atom leaves the cavity the state of the
system will be given by

\begin{eqnarray*}
\widehat{\rho }\left( 2\pi /\Omega ^{\prime }\right) &=&\widehat{\rho }%
_{gg}\left( 2\pi /\Omega ^{\prime }\right) \otimes \left| g\right\rangle
\left\langle g\right| +\widehat{\rho }_{ii}\left( 2\pi /\Omega ^{\prime
}\right) \otimes \left| i\right\rangle \left\langle i\right| \\
&&+\left[ \widehat{\rho }_{ig}\left( 2\pi /\Omega ^{\prime }\right) \otimes
\left| i\right\rangle \left\langle g\right| +{\rm h.\;c.}\right] .
\end{eqnarray*}

\noindent After the second Ramsey zone, the atomic state will once more be
changed according to (\ref{g-pi/2}), and the state of the composite system
becomes

\begin{eqnarray}
\widehat{\rho }^{^{\prime }}\left( 2\pi /\Omega ^{\prime }\right) &=&\frac{1%
}{2}\left[ \widehat{\rho }_{ii}\left( 2\pi /\Omega ^{\prime }\right) +%
\widehat{\rho }_{gg}\left( 2\pi /\Omega ^{\prime }\right) +\widehat{\rho }%
_{ig}\left( 2\pi /\Omega ^{\prime }\right) +\widehat{\rho }_{gi}\left( 2\pi
/\Omega ^{\prime }\right) \right] \otimes \left| i\right\rangle \left\langle
i\right|  \label{rho-3} \\
&&+\frac{1}{2}\left[ \widehat{\rho }_{ii}\left( 2\pi /\Omega ^{\prime
}\right) +\widehat{\rho }_{gg}\left( 2\pi /\Omega ^{\prime }\right) -%
\widehat{\rho }_{ig}\left( 2\pi /\Omega ^{\prime }\right) -\widehat{\rho }%
_{gi}\left( 2\pi /\Omega ^{\prime }\right) \right] \otimes \left|
g\right\rangle \left\langle g\right|  \nonumber \\
&&+\frac{1}{2}\left[ \widehat{\rho }_{gg}\left( 2\pi /\Omega ^{\prime
}\right) -\widehat{\rho }_{ii}\left( 2\pi /\Omega ^{\prime }\right) +%
\widehat{\rho }_{ig}\left( 2\pi /\Omega ^{\prime }\right) -\widehat{\rho }%
_{gi}\left( 2\pi /\Omega ^{\prime }\right) \right] \otimes \left|
i\right\rangle \left\langle g\right|  \nonumber \\
&&+\frac{1}{2}\left[ \widehat{\rho }_{gg}\left( 2\pi /\Omega ^{\prime
}\right) -\widehat{\rho }_{ii}\left( 2\pi /\Omega ^{\prime }\right) -%
\widehat{\rho }_{ig}\left( 2\pi /\Omega ^{\prime }\right) +\widehat{\rho }%
_{gi}\left( 2\pi /\Omega ^{\prime }\right) \right] \otimes \left|
g\right\rangle \left\langle i\right| .  \nonumber
\end{eqnarray}

\noindent At this moment, taking the limit of vanishing dissipation ($%
k\rightarrow 0$), the atomic state $i$($g$) is completely correlated with
the 0(1)-photon state of the field in the high-Q cavity. In the subspace
spanned by the vectors $\left\{ \left| 0\right\rangle ,\left| 1\right\rangle
\right\} $, the operadors $\widehat{\rho }_{ii}\left( 2\pi /\Omega ^{\prime
}\right) $, $\widehat{\rho }_{gg}\left( 2\pi /\Omega ^{\prime }\right) $ and 
$\widehat{\rho }_{ig}\left( 2\pi /\Omega ^{\prime }\right) =\widehat{\rho }%
_{gi}^{\dagger }\left( 2\pi /\Omega ^{\prime }\right) $ are given by the
matrices

\[
\widehat{\rho }_{ii}\left( 2\pi /\Omega ^{\prime }\right) =\frac{1}{2}\left( 
\begin{array}{cc}
P_{1}e^{-4\pi k/\Omega ^{\prime }} & 0 \\ 
0 & 1-P_{1}e^{-4\pi k/\Omega ^{\prime }}
\end{array}
\right) , 
\]

\[
\widehat{\rho }_{gg}\left( 2\pi /\Omega ^{\prime }\right) =\frac{1}{2}\left( 
\begin{array}{cc}
P_{1}e^{-2\pi k/\Omega ^{\prime }} & 0 \\ 
0 & 1-P_{1}e^{-2\pi k/\Omega ^{\prime }}
\end{array}
\right) , 
\]

\noindent and

\[
\widehat{\rho }_{gi}\left( 2\pi /\Omega ^{\prime }\right) =\frac{1}{2}\left( 
\begin{array}{cc}
-P_{1}e^{-3\pi k/\Omega ^{\prime }} & 0 \\ 
0 & 1-P_{1}+P_{1}\left( e^{-3\pi k/\Omega ^{\prime }}+1\right) \frac{8k^{2}}{%
9k^{2}+\Omega ^{\prime 2}}
\end{array}
\right) . 
\]

If the atomic state is {\it not} measured, the state of the field becomes

\begin{eqnarray*}
\widehat{\rho }_{F}^{^{\prime }}\left( 2\pi /\Omega ^{\prime }\right) &=&%
\mathop{\rm tr}\nolimits_{A}\widehat{\rho }^{^{\prime }}\left( 2\pi /\Omega
^{\prime }\right) \\
&=&\widehat{\rho }_{ii}\left( 2\pi /\Omega ^{\prime }\right) +\widehat{\rho }%
_{gg}\left( 2\pi /\Omega ^{\prime }\right) \\
&=&\frac{P_{1}}{2}\left( e^{-4\pi k/\Omega ^{\prime }}+e^{-2\pi k/\Omega
^{\prime }}\right) \left| 1\right\rangle \left\langle 1\right| \\
&&+\left[ 1-\frac{P_{1}}{2}\left( e^{-4\pi k/\Omega ^{\prime }}+e^{-2\pi
k/\Omega ^{\prime }}\right) \right] \left| 0\right\rangle \left\langle
0\right| .
\end{eqnarray*}

\noindent Note that this state is different from the one that would have
been obtained if the cavity field had been left to evolve undisturbed. In
this case the state of the field would have been

\[
\widehat{\rho }_{F}\left( 2\pi /\Omega ^{\prime }\right) =P_{1}e^{-4\pi
k/\Omega ^{\prime }}\left| 1\right\rangle \left\langle 1\right| +\left(
1-P_{1}e^{-4\pi k/\Omega ^{\prime }}\right) \left| 0\right\rangle
\left\langle 0\right| . 
\]

The probability for detecting the atom in state $g$ when the state of the
composite system is given by (\ref{rho-3}) is found to be

\begin{equation}
P_{g}=\frac{P_{1}}{2}\left( \frac{k^{2}+\Omega ^{\prime 2}}{9k^{2}+\Omega
^{\prime 2}}\right) \left( e^{-3\pi k/\Omega ^{\prime }}+1\right) .
\label{Pg-1}
\end{equation}

\noindent If the atom is in fact detected in state $g$, the state of the
field will be reduced to

\begin{eqnarray*}
\widehat{\rho }_{F}^{^{\prime }}\left( 2\pi /\Omega ^{\prime },g\right) &=&%
\frac{\left\langle g\right| \widehat{\rho }^{^{\prime }}\left( 2\pi /\Omega
^{\prime }\right) \left| g\right\rangle }{P_{g}} \\
&=&\frac{1}{2P_{g}}\left[ \widehat{\rho }_{ii}\left( 2\pi /\Omega ^{\prime
}\right) +\widehat{\rho }_{gg}\left( 2\pi /\Omega ^{\prime }\right) -%
\widehat{\rho }_{ig}\left( 2\pi /\Omega ^{\prime }\right) -\widehat{\rho }%
_{gi}\left( 2\pi /\Omega ^{\prime }\right) \right] \\
&=&P_{1}\left( g\right) \left| 1\right\rangle \left\langle 1\right|
+P_{0}\left( g\right) \left| 0\right\rangle \left\langle 0\right| \text{,}
\end{eqnarray*}

\noindent where we defined

\[
P_{1}\left( g\right) \equiv \frac{P_{1}}{4P_{g}}e^{-2\pi k/\Omega ^{\prime
}}\left( e^{-\pi k/\Omega ^{\prime }}+1\right) ^{2} 
\]

\noindent with $P_{1}\left( g\right) +P_{0}\left( g\right) =1$.

\vspace{.3cm}

\noindent {\bf Changing the field decay.} Assume that a one-photon state is
created in the microwave cavity and left to evolve there for a short time
interval $\Delta t$. The state of the field after this time lapse will be

\begin{equation}
\widehat{\rho }_{F}\left( \Delta t\right) =e^{-2k\Delta t}\left|
1\right\rangle \left\langle 1\right| +\left( 1-e^{-2k\Delta t}\right) \left|
0\right\rangle \left\langle 0\right| .
\end{equation}

\noindent Let next an atom be sent through the apparatus to probe the state
of the field. Call this atom the ``measuring atom''. The probability that
the measuring atom is detected in state $g$ can be calculated from (\ref
{Pg-1}) to be

\[
P_{g}\left( \Delta t\right) =\frac{e^{-2k\Delta t}}{2}\left( \frac{%
k^{2}+\Omega ^{\prime 2}}{9k^{2}+\Omega ^{\prime 2}}\right) \left( e^{-3\pi
k/\Omega ^{\prime }}+1\right) . 
\]

\noindent Consider, furthermore, the case in which a second atom (to be
called the ``monitoring atom'') be sent through the apparatus {\it during}
the time interval $\Delta t$, and let its final state remain undetected. The
state of the field after the time interval $\Delta t\geq 2\pi /\Omega
^{\prime }$ will be given, as a result of the monitoring, by

\begin{eqnarray*}
\widehat{\rho }_{F}^{(M)}\left( \Delta t\right) &=&\frac{1}{2}e^{2\pi
k/\Omega ^{\prime }}e^{-2k\Delta t}\left( e^{-2\pi k/\Omega ^{\prime
}}+1\right) \left| 1\right\rangle \left\langle 1\right| \\
&&+\left[ 1-\frac{1}{2}e^{2\pi k/\Omega ^{\prime }}e^{-2k\Delta t}\left(
e^{-2\pi k/\Omega ^{\prime }}+1\right) \right] \left| 0\right\rangle
\left\langle 0\right| .
\end{eqnarray*}

\noindent A second measuring atom is next sent in. The probability that it
is detected in state $g$ is

\[
P_{g}^{(M)}\left( \Delta t\right) =\frac{P_{1}^{(M)}\left( \Delta t\right) }{%
2}\left( \frac{k^{2}+\Omega ^{\prime 2}}{9k^{2}+\Omega ^{\prime 2}}\right)
\left( e^{-3\pi k/\Omega ^{\prime }}+1\right) . 
\]

\noindent Since

\[
P_{1}^{(M)}\left( \Delta t\right) =\frac{1}{2}e^{2\pi k/\Omega ^{\prime
}}e^{-2k\Delta t}\left( e^{-2\pi k/\Omega ^{\prime }}+1\right) , 
\]

\noindent we get

\[
P_{g}^{(M)}\left( \Delta t\right) =\frac{1}{4}e^{2\pi k/\Omega ^{\prime
}}e^{-2k\Delta t}\left( e^{-2\pi k/\Omega ^{\prime }}+1\right) \left( \frac{%
k^{2}+\Omega ^{\prime 2}}{9k^{2}+\Omega ^{\prime 2}}\right) \left( e^{-3\pi
k/\Omega ^{\prime }}+1\right) . 
\]

\noindent Taking the ratio

\begin{equation}
\frac{P_{g}^{(M)}\left( \Delta t\right) }{P_{g}\left( \Delta t\right) }=%
\frac{1}{2}\left( 1+e^{2\pi k/\Omega ^{\prime }}\right) ,  \label{ratio}
\end{equation}

\noindent we see that the probability of detecting the second measuring atom
in state $g$ is increased as a result of the monitoring process. As already
mentioned, this enhancement can be understood as resulting from the
temporary removal of the photon from the cavity by the monitoring atom, thus
making it unavailable for the decay process. An effective absence time $\tau 
$ can be defined in terms of the ratio (\ref{ratio}) as

\[
\frac{1}{2}\left( 1+e^{2\pi k/\Omega ^{\prime }}\right) =e^{\Omega ^{\prime
}\tau /2\pi } 
\]

\noindent so that clearly $0<\tau <2\pi /\Omega ^{\prime }$. If the
interaction time $2\pi /\Omega ^{\prime }$ is very short in the scale of the
decay time $1/k$, the effect of the monitoring atom on the ratio (\ref{ratio}%
) tends to disappear.

For the experimental values of $\Omega $ and $k$ in refs. \cite
{nogues,nogues2}, we get an enhancement of 0.5\% for the probability $%
P_{g}^{(M)}\left( \Delta t\right) $. This effect would hardly be detectable.
However, for lower-Q cavities, the situation may change drastically: if one
uses $10^{4}\ {\rm s}^{-1}\lesssim k\lesssim 10^{5}\ {\rm s}^{-1}$, the
renormalized Rabi frequency $\Omega ^{\prime }$\ remains essentialy
unaltered and the ratio (\ref{ratio}) varies within a large range: 
\[
1.1\lesssim \frac{P_{g}^{(M)}\left( \Delta t\right) }{P_{g}\left( \Delta
t\right) }\lesssim 4. 
\]

Recently, an experiment using a similar scheme was realized at the Ecole
Normale Sup\'{e}rieure, in Paris\cite{nogues2}. Although this experiment was
intended to measure the Wigner function at the origin of the phase space for
single photon and vacuum states, it can be adapted to verify the results
obtained here.

\end{document}